\documentclass[prl,10pt,twocolumn]{revtex4}

\usepackage{amsmath}
\usepackage{amsfonts}
\usepackage{amssymb}
\usepackage{dcolumn}
\usepackage[cp1251]{inputenc}
\usepackage[english]{babel}
\usepackage{epsfig}
\usepackage{graphics}
\usepackage{color}

\begin{document}

\title{Quasi-Coulomb series in a two-dimensional three-body system}

\author{Maxim A.\ Efremov$^{1,2}$ and Wolfgang P.\ Schleich$^{1,3}$}
\affiliation{$^1$Institut f\"ur Quantenphysik and Center for
Integrated Quantum Science and Technology ($\it IQ^{ST}$),
Universit\"at Ulm, 89081 Ulm, Germany \\
$^2$A.M. Prokhorov General Physics Institute,
Russian Academy of Sciences, 119991 Moscow, Russia \\
$^3$Texas A$\&$M University Institute for Advanced Study (TIAS), Institute for Quantum Science and Engineering (IQSE) and
Department of Physics and Astronomy, Texas A$\&$M University, College Station, Texas 77843-4242, USA}


\begin{abstract}

We show that the bound states in a three-body system display a Coulomb series with a Gaussian cut-off provided: 
(i) the system consists of a light particle and two heavy bosonic ones, 
(ii) the heavy-light short-range potential has a resonance in the $p$-wave scattering amplitude, and (iii) all three particles move in two space dimensions. 
For a decreasing mass ratio this quasi-Coulomb series merges into a pure Coulombic one.

\end{abstract}


\maketitle

\noindent{\it Introduction.--} 
Many physical phenomena depend crucially on the number of dimensions of space and two dimensions play a very special role. 
Three examples may illustrate this point. In two dimensions (i) the Huygens principle familiar from classical optics fails \cite{Huygens}, 
(ii) in quantum mechanics already an infinitesimally small binding potential suffices to create a bound state \cite{Simon},
and (iii) the Berezinskii-Kosterlitz-Thouless transition \cite{BKT} takes place only there. 
In the realm of few-body physics, the Efimov effect \cite{Efimov}, 
which is the formation of an infinite series of three-body bound states induced by a two-body $s$-wave resonance, 
only exists in three dimensions \cite{Efimov-two dimensions}. 
However, in the present Letter we show that even in two dimensions there exists an infinite series of three-body bound states 
provided the two-body interaction has a $p$-wave resonance.     

The bound states of three particles associated with the Efimov effect 
result from an effective potential, which decays with the {\it square} of the distance 
\cite{Efimov, Jensen, Nielsen-Fedorov, Braaten-Hammer} between the particles. 
In the case of a $p$-wave resonance, an universal series of three-body bound states is formed due to an effective potential, 
which decays \cite{Efremov-three-dimensions, Zhu Tan} with the {\it cube} of the distance. 
Consequently, in the Efimov effect the energy spectrum depends exponentially \cite{Efimov, Ferlaino-review, He-experiments} 
on the quantum number $n$, whereas in the case of the $p$-wave resonance it scales with the sixth power \cite{Efremov-three-dimensions}. 

We emphasize that the shape of the effective potentials and the associated energy spectra 
are intimately connected to the fact that the particles experience a three-dimensional space.
In this Letter we confine a three-body system composed of a light
atom of mass $m$ and two heavy bosonic ones of mass $M$ with a $p$-wave resonance in the heavy-light interaction 
to two space dimensions \cite{super-Efimov-comment, super-Efimov}.
We derive an effective potential, which consists of the familiar Efimov potential screened by the inverse of a logarithm. 
The length scale of the logarithm is determined by the parameters of the two-dimensional $p$-wave resonance. 

The appearance of the logarithm is a consequence of the dimensionality of the problem. 
Indeed, it originates from the asymptotic behavior of the 
two-dimensional scattering amplitude at low incident energy \cite{LL, two-dimensional scattering theory} given by a Hankel function, 
which displays a logarithmic singularity at the origin \cite{Abramowitz}.

The Langer transformation \cite{Langer} applied to the two-dimensional radial Schr\"odinger equation 
containing the product of the Efimov potential and the inverse of the logarithm gives rise to 
an effective one-dimensional Coulomb potential \cite{Loudon} with the familiar discrete Coulomb energy spectrum. 
Due to this transformation the energy eigenvalue is multiplied by an exponential and we arrive at the universal energy spectrum 
\begin{equation}
 \label{main result}
    E_n=-\frac{E_0}{n^2}\exp\left[-\frac{\pi^2}{2}\frac{\mu}{M}\,n^2\right]
\end{equation}
for large $n$. The familiar Coulomb series is modified by a Gaussian cut-off governed by the mass ratio $\mu/M$, 
where $\mu\equiv 2mM/(m+2M)$ denotes the reduced mass and $E_0$ is the characteristic energy determined
by the short-range physics. 

Similar to the Efimov effect \cite{Braaten-Hammer}, we can detect these three-body bound states 
using an atomic mixture with a large mass ratio. 
Indeed, mixtures with a $p$-wave resonance have already
been realized with ${\rm K}$ and ${\rm Rb}$
\cite{P-wave-mixture-KRb}, as well as with ${\rm Li}$ and ${\rm Rb}$
\cite{P-wave-mixture-LiRb}, corresponding to the mass ratios $m_{\rm K}/M_{\rm Rb}\approx 0.5$ and $m_{\rm Li}/M_{\rm Rb}\approx 0.1$.

\noindent{\it Binding energy for light particle.--}
Since our three-body system consists of a light particle which interacts with two heavy
particles, we employ \cite{Fonseca, atomic-molecule, Efremov-three-dimensions, Zhu Tan} the Born-Oppenheimer approximation \cite{LL}. 
As in the case of three dimensions \cite{Efremov-three-dimensions}, 
the energy spectrum of this system is determined by the relative motion of the two heavy particles 
dictated by the Schr\"odinger equation
\begin{equation}
 \label{Schrodinger heavy}
    \left\{\Delta_{\bf R}^{(2)}+\frac{M}{\hbar^2}\left[E-{\mathcal V}({\bf R})\right]\right\}\chi({\bf R})=0.
\end{equation}
Here ${\bf R}$ and $E$ denote the separation between the two heavy particles and the total three-body energy, respectively, and 
$\Delta_{\bf R}^{(2)}$ is the Laplacian in two dimensions. 

The effective interaction potential
\begin{equation}
 \label{V-eff}
    {\mathcal V}({\bf R})\equiv -\frac{[\hbar\kappa({\bf R)}]^2}{2\mu}
\end{equation}
is the bound-state energy of the light particle interacting with two non-moving heavy particles \cite{our comment}. 

The values of $\kappa=\kappa(R)$ follow \cite{appendices, Efremov-three-dimensions} from 
the condition that the determinant of the system of linear algebraic equations
\begin{eqnarray}
 \label{system-final1}
    C_m^{(+)}+i\pi\,T_m(i\kappa)\sum_{m'=-\infty}^{\infty}H_{m-m'}^{(1)}(i\kappa R)C_{m'}^{(-)}=0 \\ \label{system-final2}
    C_m^{(-)}+i\pi\,T_m(i\kappa)\sum_{m'=-\infty}^{\infty}H_{m'-m}^{(1)}(i\kappa R)C_{m'}^{(+)}=0
\end{eqnarray}
for the coefficients $C_m^{(\pm)}$ vanishes, which provides us with a transcendental equation.
Here $m=0,\pm 1,\pm 2$,... and $H_m^{(1)}$ are the Hankel functions \cite{Abramowitz} with the on-shell $T$-matrix elements \cite{Newton}
\begin{equation}
 \label{S-matrix-phase}
  T_m(i\kappa)=-\frac{1}{\pi}\frac{1}{\cot[\delta_m(i\kappa)]-i}
\end{equation}
of the heavy-light potential $U$ determined by the scattering phase $\delta_m$.
For the sake of simplicity $U$ is assumed to be spherically symmetric and to have the finite range $r_0$, that is $U(r>r_0)=0$.

We emphasize that a similar set of equations emerges \cite{Efremov-three-dimensions} in the three-dimensional case. 
However, the main difference in two dimensions is the appearance of the Hankel functions $H_m^{(1)}$ with an integer  
rather than a half-integer index. 
This substitution is a consequence of the reduced dimensionality and can be traced back \cite{appendices} to the Green function \cite{LL} 
of a free particle in two dimensions given by $H_0^{(1)}$.

\noindent{\it Efficiency of s- and p-wave scattering.--} 
Since there is no Efimov effect in two dimensions \cite{Efimov-two dimensions, Belotti, appendices}, 
we now focus on the $p$-wave resonant state in the heavy-light potential $U$ and show
that the $s$- and $p$-wave scattering are of the same order. This feature is crucial 
for truncating the system Eqs. (\ref{system-final1})-(\ref{system-final2}).

Similar to the three-dimensional scattering, in the low-energy limit, that is for $\kappa r_0\ll 1$,
the $p$-wave scattering phase
\begin{equation}
 \label{S-matrix-low-energy}
  \cot[\delta_1(i\kappa)]\cong \frac{2}{\pi}\left[\frac{1}{a_1 \kappa^2}+\ln\left(i\kappa r_1\right)\right]
\end{equation}
is parametrized within the two-dimensional effective-range expansion \cite{effective range} by
the $p$-wave effective scattering length $a_1$ and effective range $r_1$.
Both $a_1$ and $r_1$ are non-negative parameters and for $a_1\gg r_1^2$ they are determined by the energy
\begin{equation}
 \label{energy-p-wave}
  \varepsilon_1\cong -\frac{\hbar^2}{\mu a_1\ln[a_1/(2 r_1^2)]}
\end{equation}
of the $p$-wave bound state. The latter is defined by the positive-valued pole $\kappa_1$ of the matrix element $T_1$,
given by Eq. (\ref{S-matrix-phase}) with Eq. (\ref{S-matrix-low-energy}), that is $\varepsilon_1\equiv -(\hbar\kappa_1)^2/2\mu$.

Moreover, in the case of a $p$-wave resonance, that is $a_1^{-1}=0$, 
the effective range $r_1$ has an upper bound \cite{effective range} determined by the potential range $r_0$, that is $r_1\leq \frac{1}{2} e^{\gamma}r_0$, 
with $\gamma$ being the Euler constant, for any two-dimensional short-range potential. 
The matrix element $T_1$ for the resonant channel given by Eqs. (\ref{S-matrix-phase}) and (\ref{S-matrix-low-energy}) 
is of the same order as $T_{0}$ for the non-resonant $s$-wave channel \cite{LL} following from 
\begin{equation}
 \label{zero-range}
    \cot[\delta_0(i\kappa)]\cong\frac{2}{\pi}\left[\gamma+\ln\left(\frac{i\kappa a_0}{2}\right)\right].
\end{equation}
Here $a_0$ is the two-dimensional scattering length.

\noindent{\it Effective potential for p-wave resonance.--}
Since in the low-energy limit, that is $\kappa r_0\ll 1$, we have the estimate $T_{m>1}(i\kappa)\sim (\kappa r_0)^{2m}$, 
we can confine ourselves in Eqs. (\ref{system-final1}) and (\ref{system-final2}) only to the $s$- and $p$-waves, giving rise
to a system of six algebraic equations for $C_{0}^{(\pm)}$, $C_{1}^{(\pm)}$, and $C_{-1}^{(\pm)}$. This system has non-trivial
solutions only if the corresponding determinant vanishes, which provides us with two branches of solutions \cite{appendices} for $\kappa$ 
corresponding to different wave functions of the light particle.

The first branch is determined \cite{appendices} by the transcendental equation
\begin{equation}
 \label{energy-eq-s1-first-dimensionless}
   K_0\left(\xi\,\frac{R}{r_1}\right)-K_2\left(\xi\,\frac{R}{r_1}\right)=\pm\left(\frac{r_1^2}{a_1}\frac{1}{\xi^2}+\ln\xi\right)
\end{equation}
for $\xi\equiv\kappa r_1$, where $K_m$ are modified Bessel functions \cite{Abramowitz}.

In the resonant case, that is $a_1^{-1}=0$, Eq.
(\ref{energy-eq-s1-first-dimensionless}) has a solution $\xi_1$ only
for the plus sign on the right-hand side due to the asymptotic behavior \cite{Abramowitz} $K_2(z)\cong 2/z^2$ and $K_0(z)\cong -\ln(z)$ for $z\rightarrow 0$.
In the limit of $0<\xi_1\ll 1$, we find $\xi_1^2\cong
2e^{1-2\gamma}[\rho^2\ln(\rho\ln\rho)]^{-1}$ for $\rho\equiv \exp(\frac{1}{2}-\gamma)R r_1^{-1}\gg 1$,
giving rise to the effective potential
\begin{equation}
 \label{energy-asymptotic-first}
    {\mathcal V}_{{\rm I}}^{(0)}(R)\cong
    -\frac{\hbar^2}{\mu R^2}\frac{1}{\ln\frac{R}{r_1}-\gamma+\frac{1}{2}+\ln\left(\ln\frac{R}{r_1}-\gamma+\frac{1}{2}\right)}
\end{equation}
for $R\gg r_1$. Here the subscript ${\rm I}$ refers to the first branch.

However, near the resonance, $a_1\gg r_1^2$, that is in the
case of the weakly-bound $p$-wave state in $U$, the form of the effective potentials 
$$
{\mathcal V}_{{\rm I}}^{(\pm)}\equiv -\frac{\hbar^2}{2\mu r_1^2}\left(\xi_{{\rm I}}^{(\pm)}\right)^2
$$
is determined by the two solutions $\xi_{{\rm I}}^{(\pm)}$ of Eq. (\ref{energy-eq-s1-first-dimensionless}). 
The ranges of ${\mathcal V}_{{\rm I}}^{(+)}$ and ${\mathcal V}_{{\rm I}}^{(-)}$ are identical and equal to 
\begin{equation}
 \label{range R_1}
    R_1\equiv\frac{\hbar}{\sqrt{2\mu|\varepsilon_1|}}\cong \left(\frac{a_1}{2}\ln\frac{a_1}{2r_1^2}\right)^{\frac{1}{2}}.
\end{equation}
For large distances, $R\gtrsim R_1$, they approach exponentially the bound state energy $\varepsilon_1$ of the light particle 
defined by Eq. (\ref{energy-p-wave}). For short distances, $R\lesssim R_1$, 
${\mathcal V}_{\rm I}^{(+)}$ decreases monotonically and approaches ${\mathcal V}_{\rm I}^{(0)}$, Eq.
(\ref{energy-asymptotic-first}), ${\mathcal V}_{\rm I}^{(+)}(R\lesssim R_1)\cong {\mathcal V}_{\rm I}^{(0)}(R)$, 
whereas ${\mathcal V}_{\rm I}^{(-)}$ increases monotonically and vanishes at $R=\sqrt{2a_1}$.

The second branch of solutions of the truncated system of equations, 
Eqs. (\ref{system-final1})-(\ref{system-final2}), is determined \cite{appendices} by the transcendental equation
\begin{eqnarray}
 \label{energy-eq-s1-second-dimensionless}
   \left[K_2\left(\xi\frac{R}{r_1}\right)+K_0\left(\xi\frac{R}{r_1}\right)\pm\left(\frac{r_1^2}{a_1}\frac{1}{\xi^2}+\ln\xi\right)\right]\times\nonumber \\
    \left[K_0\left(\xi\frac{R}{r_1}\right)\mp\ln\left(\xi\frac{e^{\gamma}}{2}\frac{a_0}{r_1}\right)\right]=2K_1^2\left(\xi\frac{R}{r_1}\right)
\end{eqnarray}
for $\xi\equiv \kappa r_1$.

In the resonant case, that is $a_1^{-1}=0$, Eq.
(\ref{energy-eq-s1-second-dimensionless}) has a solution $\xi_2$ only
for the plus sign on the left-hand side and, in the limit of
$0<\xi_2\ll 1$, we find $\xi_2^2\cong
2e^{1-2\gamma}\left[\rho^2(\ln\rho+2\gamma+1-\ln2)\right]^{-1}$ for $\rho\gg 1$,
giving rise to the effective potential
\begin{equation}
 \label{energy-asymptotic-second}
    {\mathcal V}_{{\rm II}}^{(0)}(R)\cong
    -\frac{\hbar^2}{\mu R^2}\frac{1}{\ln\frac{R}{2r_1}+\gamma+\frac{3}{2}}
\end{equation}
for $R\gg r_1$. Here the subscript ${\rm II}$ refers to the second branch. 
For a more detailed discussion of the potentials ${\mathcal V}_{\rm II}^{(\pm)}$ arising in the second branch, 
we refer to Ref. \cite{appendices}.

On the $p$-wave resonance and for $R\rightarrow \infty$, both potentials ${\mathcal V}_{\rm I}^{(0)}$ and ${\mathcal V}_{\rm II}^{(0)}$ 
given by Eqs. (\ref{energy-asymptotic-first}) and
(\ref{energy-asymptotic-second}) approach the same behavior
\begin{equation}
 \label{V asymptotic}
  {\mathcal V_{\rm I}^{(0)}}(R\rightarrow \infty)\cong {\mathcal V_{\rm II}^{(0)}}(R\rightarrow \infty)\cong 
  {\mathcal V}(R)\equiv -\frac{\hbar^2}{\mu R^2}\frac{1}{\ln\frac{R}{r_1}}
\end{equation}
determined by effective range $r_1$ of the $p$-wave.

\noindent{\it Energy spectrum induced by effective potential.--} 
Next we focus on the two-dimensional dynamics of the two heavy bosonic particles dictated by the
Schr\"odinger equation (\ref{Schrodinger heavy}) with the potential ${\mathcal V}$ given by Eq. (\ref{V asymptotic}). 
In particular, we show that ${\mathcal V}$ supports an infinite number of three-body bound states following a quasi-Coulomb series.

For the case of zero orbital angular momentum between the two heavy bosonic particles, we can use the WKB solution \cite{appendices} 
\begin{equation}
 \label{WKB solution}
    \chi_E(R)\propto\sin\left[\varphi(R,R_E)\right]
\end{equation}
of Eq. (\ref{Schrodinger heavy}) with the phase
\begin{equation}
 \label{WKB phase}
    \varphi(R,R_E)\equiv\frac{1}{\hbar}\int_{R}^{R_E}dR'\sqrt{M[E-{\mathcal V}(R')]}+\theta_E
\end{equation}
accumulated between $R$ and the outer turning point $R_E$ determined by the condition ${\mathcal V}(R_E)=E$. 
Here the phase $\theta_E$, with $|\theta_E|\leq \pi$, which depends only weakly on the energy $E$, is determined by the behavior of 
the potential at short distances, that is for $R\sim r_1 \sim r_0$.

Since the spectrum of the bound states in ${\mathcal V}$ has an accumulation point at $E=0$, 
we need to know the behavior of $\varphi$ as $E\rightarrow 0$. For this purpose, we neglect 
in the limit of small and negative energy, $|E|\ll \hbar^2/(M r_1^2)$, and large distances, $r_1\ll R \ll R_E$, 
the energy $E$ under the square root on the right-hand side of Eq. (\ref{WKB phase}) and obtain with Eq. (\ref{V asymptotic}) the approximation
$$
\varphi\cong \sqrt{\frac{M}{\mu}}\int_{R}^{R_E}\frac{dR'}{R'\sqrt{\ln\frac{R'}{r_1}}}+\theta_0
$$
for the accumulated phase \cite{appendices}, or
\begin{equation}
 \label{WKB phase approximate first}
    \varphi(R,R_E)\cong 2\sqrt{\frac{M}{\mu}}\left[\left(\ln\frac{R_{E}}{r_1}\right)^{\frac{1}{2}}-\left(\ln\frac{R}{r_1}\right)^{\frac{1}{2}}\right]+\theta_0.
\end{equation}

The energy spectrum of the weakly bound states induced by the potential ${\mathcal V}$ 
follows from the familiar WKB quantization rule $\varphi(r_1,R_{n})=\pi n$, giving rise to the discrete positions 
$R_{n}\cong r_1\exp[(\mu/4M)\pi^2 n^2]$ of the outer turning points for $n\gg 1$. 
The connection $E_n={\mathcal V}(R_{n})$
between the binding energy $E_n$ and $R_{n}$ finally yields the asymptotic energy spectrum, Eq. (\ref{main result}),  
in the form of the Coulomb series with a Gaussian cut-off governed by the mass ratio.
The characteristic energy $E_0\sim \hbar^2/(\mu r_0^2)$ is determined by the short-range physics of ${\mathcal V}$. 

As a result, an exact two-dimensional $p$-wave resonance in the heavy-light short-range interaction
potential creates the two long-range effective potentials ${\mathcal V}_{{\rm I}}^{(0)}$, Eq. (\ref{energy-asymptotic-first}), and 
${\mathcal V}_{{\rm II}}^{(0)}$, Eq. (\ref{energy-asymptotic-second}). They both merge into the same asymptotic potential 
${\mathcal V}$, Eq. (\ref{V asymptotic}), which gives rise to an infinite series of the weakly bound three-body states.

However, near the resonance, that is for large but finite values of $a_1$, 
the range $R_1$ given by Eq. (\ref{range R_1}) is finite and the effective potentials 
${\mathcal V}_{\rm I}^{(0)}$ and ${\mathcal V}_{\rm II}^{(0)}$ are valid only in the region $r_1\lesssim R\lesssim R_1$.
As a result, the number $N_0$ of bound states is finite and given by the number of nodes of the zero-energy solution $\chi_0$ 
defined by Eq. (\ref{WKB solution}) with the phase $\varphi(R,R_1)$, Eq. (\ref{WKB phase approximate first}), accumulated between $R$ and $R_E\sim R_1$. 
Since $R_1\gg r_1$, we can estimate $N_0\cong \varphi(r_1,R_1)/\pi$ with the help of Eq. (\ref{range R_1}) as
\begin{equation}
 \label{N bound states}
    N_0\cong\frac{2}{\pi}\left(\frac{M}{\mu}\ln\frac{R_1}{r_1}\right)^{\frac{1}{2}}\cong
    \frac{1}{\pi}\left(2\frac{M}{\mu}\ln\frac{a_1}{2r_1^2}\right)^{\frac{1}{2}}.
\end{equation}
Thus, $N_0$ increases with the square root of $M/\mu$ and diverges weakly as a logarithm of $a_1/r_1^{2}$ when we tune $a_1$ closer
to the $p$-wave resonance, that is for $a_1\rightarrow \infty$.

\noindent{\it New states as resonances in atom-molecule scattering.--} 
The appearance of the binding potentials can be verified experimentally 
by scattering a heavy atom off the diatomic molecule consisting of the heavy and the light atom \cite{atomic-molecule, Hammer-Petrov}. 
The predicted three-body bound states manifest themselves as resonances in the cross section of the
atom-molecule scattering when we tune the scattering length with Feshbach resonances \cite{Feshbach-review} and 
approach in this way the two-dimensional $p$-wave resonance. 

Indeed, at the low incident energy $(\hbar k)^2/M\equiv E-\varepsilon_1$, such as
$kr_1\ll 1$, the total atom-molecule cross-section \cite{LL}
$$
\sigma_0=\frac{\pi^2}{k}\left[\frac{\pi^2}{4}+\ln^2\left(\frac{kA_0}{2}\,e^{\gamma}\right)\right]^{-1}
$$
is determined mainly by the two-dimensional atom-molecule scattering
length $A_{0}$. For large values of $a_1$, it can be estimated by matching at $R\sim R_1$ 
the logarithmic derivative of $\chi_0(R)\propto\sin[\varphi(r_1,R)]$ with $\varphi$ for $R\lesssim R_1$, 
with the logarithmic derivative of the spherically-symmetric zero-energy solution $\chi(R)\propto\ln(R/A_0)$ 
of Eq. (\ref{Schrodinger heavy}) valid for $R\gtrsim R_1$, where the effective potential ${\mathcal V}$ vanishes.
Indeed, with the help of Eq. (\ref{WKB phase approximate first}) for $\varphi(r_1,R)$, we arrive at 
$$
\frac{1}{\ln(R_1/A_0)}=R_1\frac{\chi'_0(R_1)}{\chi_0(R_1)}=\frac{2M}{\mu}
\frac{\cot\left(\pi N_0\right)}{\pi N_0}
$$
with $N_0$ given by Eq. (\ref{N bound states}). 

Hence, the atom-molecule scattering length
$$
A_0=\left(\frac{a_1}{2}\ln\frac{a_1}{2r_1^2}\right)^{\frac{1}{2}}
\exp\left\{-\frac{\mu}{2M}\frac{\pi N_0(a_1)}{\cot\left[\pi N_0(a_1)\right]}\right\}
$$
exhibits an infinite series of resonances at
$a_1=a_1^{(n)}$, where $a_1^{(n)}$ are the zeros of the equation $N_0(a_1^{(n)})=(n+\frac{1}{2})$.
For $n\gg 1$, Eq. (\ref{N bound states}) provides with the expression 
$$
a_1^{(n)}\sim r_1^2\exp\left[\frac{\pi^2}{2}\frac{\mu}{M}n^2\right]
$$
for the position of the resonances and for a small mass-ratio many resonances are visible.

\noindent{\it Summary.--} We have found a novel series of bound
states in a three-body system consisting of a light particle and two heavy bosonic ones when the
heavy-light short-range interaction potential has a
two-dimensional $p$-wave resonance and the system is constrained to two space dimensions. In the case of an exact
resonance, the effective potentials between the two heavy particles are attractive and of
long-range and support an infinite number of bound states. 
The spectrum has the form of the Coulomb series with a Gaussian cut-off governed by the mass ratio.
We emphasize that these results are a consequence of an intricate interplay between the symmetry properties of 
the underlying resonances and the dimensionality of the problem. Throughout this Letter we have focused on 
an atomic system but we envision applications in nuclear physics as well.

\noindent{\it Acknowledgments.--} We are deeply indebted to K. Bongs and F. Ferlaino for stimulating discussions. 
This work was supported by a grant from the Ministry of Science, Research and the Arts of
Baden-W\"urttemberg (Az: 33-7533-30-10/19/2) to M.A.E. We also
appreciate the funding by the German Research Foundation (DFG) in the framework of the SFB/TRR-21. 
W.P.S. is grateful to Texas A$\&$M University for a Texas A$\&$M University Institute for Advanced Study (TIAS) Faculty Fellowship.
M.A.E. thanks the Alexander von Humboldt Stiftung and the Russian Foundation for Basic Research (10-02-00914-a).


\begin{thebibliography}{99}
\bibitem{super-Efimov-comment} Our work is different from the so-called ``super-Efimov`` effect \cite{super-Efimov} in two aspects: 
(i) large mass ratio versus identical masses, and (ii) two heavy bosons versus three polarized identical fermions.
\bibitem{our comment} In Eq. (\ref{Schrodinger heavy}) we neglect the interaction between the two heavy particles, since for atoms
it is typically short-range and has for large distances a van-der-Waals tail, $1/R^6$, 
with no effect on the total potential at large distances.
\end{thebibliography}
\end{document}